\documentclass[12pt]{iopart}
\usepackage{graphicx}
\usepackage[caption=false]{subfig}
\usepackage{epstopdf}
\usepackage{amsfonts}
    
\begin{document}

%\noindent\textsc{IARD 2020 conference paper}

   \title{Observers, observations and referencing in physics theories}

   \author{Uri Ben-Ya'acov}

   \address{School of Engineering, Kinneret Academic College on
   the Sea of Galilee, \\   D.N. Emek Ha'Yarden 15132, Israel}

   \ead{uriby@kinneret.ac.il}

%   ({\today})

%\vskip 2.0cm

\begin{abstract}

Is it possible to encompass the full extent of the universe within a theory based on a finite set of first principles and inference rules?

The r\^{o}le of observers and observations in physics theories is considered here in the light of G\"{o}del's incompleteness theorem. Physics theories are the sum-total that we -- humans, scientists, physicists -- can make in interpreting our observations of the universe. We are integral part of the universe, together with our observations, therefore acts of observation are also observables and should become part of the phenomena considered by the theory, especially in view of the fact that arbitrarily chosen modes of observations may essentially determine empirical results.

Incompleteness arises in G\"{o}del's theorem with self-referential propositions. Observations and interpretations are acts of referencing, and self-referencing occurs in physics whenever the observer is recognized as being part of the observed system. If self-reference appears in physics in simile to G\"{o}del's theorem, then incompleteness seems unavoidable in physics.

The article discusses observers and observations as referencing in physics, culminating with the understanding that they are hierarchically inter-related so that a universal physics theory cannot be complete.
\end{abstract}


\vskip30pt

\noindent\textit{``Humans are the universe's way of knowing itself''}

\hskip250pt\emph{Anonymous}

\vskip30pt

\eject

\section{Introduction}\label{sec: intro}

Is it possible to arrive at an ultimate theory of the universe ?

The belief that it is possible to arrive at a complete theory (sometimes known as ``theory of everything'' (TOE)) that fully describes the whole of the physical world -- a theory that accounts, via few and simple first principles and inference rules, for all the phenomena already observed and that will ever be observed -- has been, for many years and for most researchers, a fundamental tenet and the major drive for scientific research. A. Einstein put it very clearly :
\begin{quote}
\textit{It is the grand object of all theory to make these irreducible elements as simple and as few in number as possible, without having to renounce the adequate representation of any empirical content whatever.} \cite{Einstein}
\end{quote}
By ``these irreducible elements'' Einstein referred to the fundamental first principles of the theory.

However, is the physics that we know today (or that we may know at any given time) all that there is to know? Can we ever be sure that essentially new phenomena will never be observed any more?

We certainly can’t. History definitely tells the opposite, and we should always keep in mind that whatever physics theory we have in hand it is no more than the best sum-total that we can make in interpreting our observations. Yet the vision is very strong, and very domineering upon the scientific community.

Can we at least predict, speculate into the future, regarding the next phase of physics? Can we get hints from what is already known?

To get a glimpse into what seems currently to be beyond the horizon of physics, we start, from a universal point of view, with the understanding that while physics is the best interpretation that we can make of what we observe in the universe, we -- humans, scientists, physicists -- are integral part of the universe, together with our observations. These observations not only refer directly to physical phenomena, but also include observations of observations, and observations of observations of observations, \emph{etc.} These acts of observations should therefore be considered part of the subject matter of physics theories, especially when a complete all-encompassing whole-universal theory is asked for.

In classical physics, till the end of 19$^{\rm th}$ century, the common view was that the human-observer-scientist is a separated, not involved, by-standing witness to all universal phenomena. But 20$^{\rm th}$ century physics made us realize that in many instances the observer is capable of influencing the outcome of experiments. Therefore, also from the empirical point of view, the observer should be recognized as a full participant, an integral part of the observed system, with the acts of observation being also observables that should become part of the phenomena considered by the theory.

Observations and interpretations are acts of referencing. Whenever observers and observations are also part of the observed systems, participating in physical phenomena, with observers referring to their own observations, these are manifestations of self-reference in physics.

Self-reference is also the core principle of G\"{o}del's incompleteness theorem \cite{Godel,Hofstadter,NagelNewman,Raatikainen}, which implies that \emph{any rich enough consistent formal structure, based on a finite number of first principles and inference rules, cannot be complete}. G\"{o}del's theorem concerns arithmetics and logic, but since physics theories use mathematics and are organized as formal structures then naturally comes the question: Does G\"{o}del's theorem apply to physics?

The significance to physics is tremendous. Completeness implies the expectation that the predictions of the theory encompass all the phenomena already observed and those that will ever be observed, within the realm covered by the theory. In terms of physics theories this implies determinism -- if the necessary initial data are given then the state of a physical system can be predicted, at least in principle, any time in the future. Classical physics and relativity are deterministic. But the broader the domain a theory encompasses then the more it is apt to satisfy the conditions of G\"{o}del's theorem. Therefore, the natural candidate here is the "ultimate universal theory'' or "theory of everything" as defined above. If G\"{o}del's theorem applies to physics then the mere possibility of such a theory is seriously questionable.

In the following we propose to associate the recognition that observers and observations are also part of the observed systems, implying self-reference in physics, together with G\"{o}del's theorem-like consequences. This idea is not entirely new, but hardly found in physics literature. The only explicit mention of it which I am aware of is in a lecture given by S.W. Hawking almost two decades ago,
\begin{quote}
\textit{.. we are not angels, who view the universe from the outside. Instead, we and our models are both part of the universe we are describing. Thus a physical theory is self referencing, like in G\"{o}del’s theorem. One might therefore expect it to be either inconsistent or incomplete. The theories we have so far are both inconsistent and incomplete.} \cite{Hawking02}
\end{quote}

\vskip15pt

The applicability of G\"{o}del's incompleteness theorem to physics was initially discussed along these lines in \cite{GodelUPT}. In the present article we elaborate on these ideas and further the study of the r\^{o}le of referencing and self-referencing in physics.

We start by discussing referencing and self-referencing in the light of G\"{o}del's theorem, then in physics theories formed from our observation of the universe.

The questions that the article follows are ``How to describe referencing in physics? Does self-reference lead to incompleteness in physics?''

It is argued that self-reference is found in physics in observations of observations. Moreover -- in many instances the same physical phenomenon may be viewed in more than one way, so that the mode of the observation, freely chosen by the observer, determines its consequences. Self-reference and referencing in observations lead to identifying levels of observation, between observer and observations. Each level of observation requires a higher one for the former to be observed, thus creating a (potentially infinite) hierarchy of levels of observation.

Observations are followed by interpretations, therefore each higher level suggests a more profound insight which empirically implies an essentially new discovery, and together a potentially infinite hierarchy of levels of interpretation. New discoveries imply new first principles in the foundation of the theory, so the theory remains open and cannot be complete.

\vskip30pt

\section{Can G\"{o}del's theorem be applicable to physics?}\label{sec: aa}

Physics theories are expected to be complete formal (logical) systems, with a finite, consistent set of fundamental principles (axioms) and a set of deduction and inference rules which together produce, with the power of logic, predictions regarding natural phenomena. But a fundamental property of logic embodied in G\"{o}del's incompleteness theorem \cite{Godel,Hofstadter,NagelNewman,Raatikainen} implies that
\begin{quote}
\textit{In any consistent and rich enough formal structure, based on a finite number of first principles and inference rules, there will always be propositions that may be formulated within this formal system but are undecidable. Such a theory cannot be both consistent and complete.}
\end{quote}

Strictly speaking, G\"{o}del’s theorem applies to arithmetics and logic. But since physics uses mathematics, which is based on arithmetics, and logic, it seems quite natural to ask whether the results of G\"{o}del’s theorem apply to physics.

The crux of G\"{o}del's proof is in the ability to formulate, in arithmetic terms, a meta-arithmetic formula $\mathcal{G}$ that negates itself, practically defying any possible attempt to prove or refute it. The self-negation in $\mathcal{G}$ is reminiscent of the self-negation in logical paradoxes such as the liar paradox \cite{LiarPar} or Russel's paradox \cite{RussellPar}. $\mathcal{G}$ is then an undecidable arithmetic formula, demonstrating that arithmetics is incomplete.

So far physics theories do not have such a metaphysical view that allows formulation of metaphysical statements in ordinary physical terms; then it would have been possible to directly examine the applicability of G\"{o}del’s theorem to physics. However, we may proceed in a more subtle way, identifying the core principe of G\"{o}del’s theorem and examine in what way it may appear in physics.

Early attempts in favour of applying G\"{o}del's theorem to physics argued that ``G\"{o}del's theorem applies to arithmetics which is the basis of mathematics, physics uses mathematics as its language, therefore G\"{o}del's theorem applies to physics'' \cite{Jaki04,Jaki06}. The counter-argument immediately raised claimed that while G\"{o}del’s theorem is about arithmetics, there are math branches to which it does not apply (\emph{e.g.}, geometry, analysis) \cite{Raatikainen}; since these are the types of math that physics uses, there’s no reason to expect that G\"{o}del's theorem must apply to physics \cite{Barrow,Feferman06,Robertson00}. Still, the counter-argument doesn't preclude the possibility that G\"{o}del's theorem applies to physics by other arguments.

The mathematical argument is therefore non-conclusive. We have to look at the issue from a different direction.

Close inspection of G\"{o}del's theorem reveals that the core principle it relies upon is \emph{self-reference}\cite{Bolander}: If self-reference is possible then \emph{there can always be well formulated propositions that are undecidable}. These undecidable propositions are characterized by self-negation, appearing when the system asks to define itself negatively in its own terms (in other words -- refute itself), leading to logical conflicts and paradoxes in a manner similar to the liar or Russel's paradoxes. Not all self-referencing is paradoxical, and not all negation is paradoxical, but self-referencing allows paradoxical self-negation.

The present article proposes that self-reference is the decisive component of G\"{o}del's theorem where physics is concerned. Focusing on referencing and self-reference liberates us from the mathematical argument -- it is not important any more what kinds of mathematics are used by physics.

\vskip15pt

In the basis of the scientific research is the expectation that it is possible to identify in the totality of observations common fundamental principles that can be grasped by human cognition. These fundamental principles form – as axioms – the basis of the theory, and from them are deduced properties, statements and conclusions corresponding to the object of the research. Then it is expected that the combination of logical inferences with the results of observations makes it possible to examine the correctness of these fundamental principles. But G\"{o}del's theorem challenges the co-existence of the two essential characteristics that are expected from any physics theory -- consistency and completeness, both logically and physically:
\begin{itemize}
  \item Logical consistency -- that the theory does not produce conflicting predictions.
  \item Physical consistency -- that the theory does not produce predictions that contradict physical observations.
  \item Logical completeness -- that all the predictions of the theory are uniquely concludable.
  \item Physical completeness -- that with given initial data, the future can be predicted with any desired accuracy.
\end{itemize}
The last feature is the requirement of determinism.

In G\"{o}del's theorem self-reference allows paradoxical self-negation, that implies undecidability which leads to incompleteness. Accordingly, with self-reference being possible in physics, incompleteness seems unavoidable in physics theories that are large enough. In this way G\"{o}del's theorem casts doubt on the possibility of the existence together of these characteristics for sufficiently broad physics theories.

\vskip30pt

\section{Referencing and self-reference}\label{sec: refself}

Referencing and self-reference have been discussed intensively in the context of logic (see, e.g., \cite{Bolander} and references therein) and art (as in \Fref{fig:hands} below), while hardly considered in the context of physics. In the present section we consider referencing and self-reference \emph{per-se}, in a way appropriate to extend to the presentation of referencing and self-reference in physics in the following sections.

G\"{o}del's formula $\mathcal{G}$, the crux of G\"{o}del's proof, is an arithmetic formula that says ``$\mathcal{G}$ cannot be proven''. Such a definition may look very odd, due to its circularity, but G\"{o}del managed to cast it in precise arithmetic context, so it can get a numerical value. The sentence ``$\mathcal{G}$ cannot be proven'' is \textit{about} arithmetics (since $\mathcal{G}$ is an arithmetic formula) so $\mathcal{G}$ is also a \textit{meta-arithmetic} formula, and since it refers back to $\mathcal{G}$ it is self-referential. The self-negation in $\mathcal{G}$ is reminiscent of the self-negation in logical paradoxes such as the liar or Russel's paradoxes, and similarly $\mathcal{G}$ is an undecidable arithmetic formula, demonstrating that arithmetics is incomplete.

Self-negation is possible when referencing and self-referencing are possible. Let us use an analogy to illustrate and clarify \textit{referencing} vs. \textit{self-referencing}. Imagine a group of children playing in the yard. Then an adult calls them from a balcony, which is some meters above the yard. The height difference puts the balcony position at superiority relative to the yard level. If arithmetic statements are like the children playing in the yard, then meta-arithmetic statements are like the adult calling from the balcony. This is \textit{reference}. G\"{o}del's theorem deals with the meta-arithmetic statement in arithmetic terms, which is like the balcony being in the yard's level, so that both children and adult may now refer to each other on the same footing. This is \textit{self-reference}.

Referencing and self-referencing may be formally represented as follows. The analogy above suggests introducing the concept of \textit{reference levels}, with the relative status of \emph{referrer level} and \emph{referent level}. Let $a$ denote the referent and $b$ denote the referrer, with a unidirectional reference relation $b \searrow a$ between them.

For instance, $a$ may be some statement $\mathcal{S}$ while $b$ is a statement about $\mathcal{S}$, as in G\"{o}del's theorem; or, ``$b$ declares attribute $a$'' (as in the liar paradox) or ``$a$ is/isn't a member of set $b$'' (as in Russel's paradox). Referencing may also be demonstrated graphically, as in works of the Dutch artist Escher\footnotemark[1], in particular ``Drawing hands'' (\Fref{fig:hands}) with ``$b$ draws figure $a$''.

\footnotetext[1]{M.C. Escher. His works can be found on the official website http://www.mcescher.com/ and on a wide variety of other websites.}

\begin{figure}[h]
  \begin{minipage}[b]{200pt}
  \caption{``Drawing Hands''\\M.C. Escher (1948)}\label{fig:hands}
  \end{minipage}
  \hskip10pt
  \includegraphics[width=5cm]{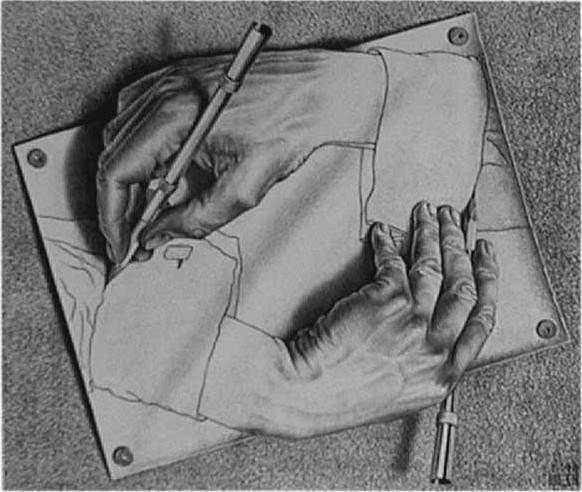}
\end{figure}

If the two levels -- referrer level and referent level -- are distinct, than the referrer level may be regarded as \textit{higher} or \textit{superior} (`balcony') to the referent level (`yard'). This is just referencing. But if referrer and referent levels are not distinguished then referencing is possible in both directions, $b \searrow a$ and $a \searrow b$, and that is self-referencing. Then paradoxical self-negation is possible, as manifested in the above examples.

It is important to note that self-reference is a necessary condition for logical paradoxes, but not sufficient. The children and the adult being in the same (yard) level doesn't necessarily mean that there's a conflict. But if the adult calls the children to end their playing and come home for dinner then a conflict is likely to occur. Similarly, referring to Russel's paradox, we may create a set $N$ which is ``the set of all sets that are members of themselves". There is no paradox here. But if such a construction is allowed, then negation may also be introduced and Russel's paradox ensues.

Consider, as another example, a space, or realm, of propositions $\left\{p\right\}$, and the doublets:

\begin{enumerate}
  \item $p \equiv$ `$q$ is true' \qquad $q \equiv$ `$p$ is true'
  \item $p \equiv$ `$q$ is false' \qquad $q \equiv$ `$p$ is false'
  \item $p \equiv$ `$q$ is true' \qquad $q \equiv$ `$p$ is false'
\end{enumerate}
The building blocks (`$p$ is true', {\textit etc.}) are perfect logical statements in all three cases, and each combination is self-referential. The first two are tautologies, even though (ii) contains negation, while (iii) is paradoxical negation.

It follows, therefore, that when we scan the whole spectrum of possibilities allowed by self-referencing, self-negation may be found there as part of the possibilities -- if affirmation is possible, so is negation.

\vskip30pt

\section{Participating observers and self-referencing}\label{sec: Parobs}

Any physics theory (or, more broadly, natural science theory) may be viewed as the conclusion and summing-up of our interpretations of our observations of nature. In an act of observation the observer refers to the observed phenomenon, hence observations are referencing in physics. We may observe our own observations, thus any observation is also an observable phenomenon. Therefore, the act of observation is referencing in physics, and self-observation is self-referencing in physics.

Classical physics is dominated by the view expressed by Victor Hugo who once said that \emph{Creation lives and evolves; the human is only a witness}. This was the common view up until the 20$^{\rm th}$ century. Now it becomes evident that we humans are not simply bystanders on the cosmic stage -- we are active participants in the evolution of the universe, the cosmos being made real in part by our own participations. This is the viewpoint put forward by people like E. Wigner and J.A. Wheeler, referring to observations of quantum phenomena. It is certainly reminiscent of the ancient Jewish tradition, that the human is active participant in Creation.

The principle of relativity manifests the fact that measurements are observer-dependent. In quantum experiments the way the experiment has been set up and the chosen mode of observation may determine the outcome of the experiment -- the very nature of the empirical end-result  -- whether the observed object is detected as a wave or a particle, or which path it follows in traveling from one point to another, \emph{etc.}.

Then the observer is not a by-stander, uninvolved, separated witness of physical phenomena but an active participant involved in the physical occurrence. The mode of observation -- arbitrarily chosen by the observer -- determines, even in small, the way the universe evolves.

\begin{figure}[h]
  \begin{minipage}[b]{200pt}
  \caption{Self-reference in physics: the observer, the observation and the subject are all part of physical phenomena}\label{fig:supersystem}
  \end{minipage}
  \hskip10pt
  \includegraphics[width=8cm]{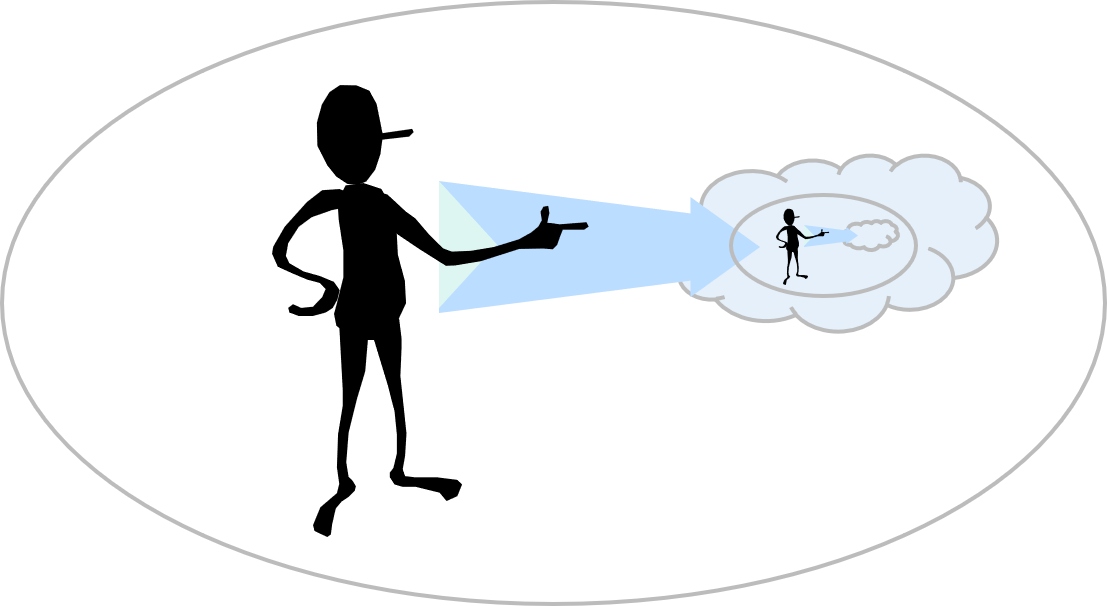}
\end{figure}

When the act of observation, being influential in physical phenomena, becomes an integral part of the phenomenon, then with it, necessarily, also the human observers, which become participating observers. When the observer, the observation and the subject are all part of the physical phenomena, this is a manifestation of self-reference in physics (\Fref{fig:supersystem}).

Summing up, physics is our (human) interpretation of what we observe in the universe. We (humans) are part of the universe, therefore we are both observers and observed. If the physical process is independent of the observer then the observer seems to be only a witness. But if the observer intervenes with the physical process then the act of observation should be considered an active part of physical processes, therefore part of the subject matter of physics theories. Observers are therefore active participants in the physical happening.

In the following sections we explore referencing observations, first in principle and then in current physics.

\vskip30pt

\section{Referencing observations}\label{sec: refobs}

In arithmetics, G\"{o}del's theorem demonstrates that the possibility of self-referencing -- arithmetic statements that are also meta-arithmetic -- allows paradoxical self-negation which in turn implies undecidability, and therefore incompleteness. In physics, so far, we don't have such physical-metaphysical statements or formulae. However, we observe the universe, then construct physics theories from the interpretation of our observations. These observations and interpretations are like the referencing in G\"{o}del's theorem.

In an observation some input enters into our mind and consciousness\footnote{Mind and consciousness are meant here, in a broad sense, as the domain where mental processes take place and we interpret our experiences, whether internal or external, and find meaning and significance for them.}. We interpret the observation referring to the image of the subject of the observation in our mind. No matter which devices are used for measurement and registration, at the end it is human interpretation. Science is the end result that we make of these interpretations.

Phenomena may be observed from a multitude of points of view. Also, possibly via various modes of observation, depending upon the way we focus our attention while observing and upon the choice of the detecting device. To illustrate this understanding we start with some graphical examples.

\begin{figure}[h]
  \begin{minipage}[b]{240pt}
  \caption{Rubin's vase: Cup or faces?}\label{fig:cupfaces}
  \end{minipage}
  \hskip10pt
  \includegraphics[width=5cm]{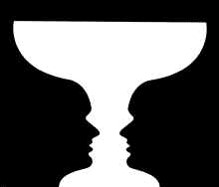}
\end{figure}

Consider first the drawing known as ``Rubin's vase'' (\Fref{fig:cupfaces}). The first thing to notice and emphasize is the fact that the same drawing may be interpreted in more than one way, the mode of interpretation depending entirely upon our choice: The drawing itself is just raw data -- a set of pixels; when these pixels enter our mind, their interpretation depends on what we choose to focus upon -- the black or the white area.

\noindent If we focus our mind on the white area we see a cup; but if we focus on the black area we see two faces. We can even switch our attention from the black to the white area, and back, thus skipping between cup and faces. Here the drawing is the subject, or referent, and the state of mind which focuses on either the white or black areas is the referrer level. But then, there is also a higher state of mind that monitors our observations of the drawing and appreciates both options. This higher state of mind is a referrer level for the state of mind which sees either the cup or the faces, which in this relation becomes the referent.

Similarly, consider Escher's ``Drawing hands'' (\Fref{fig:hands}). In first-level observations we see (focus on) either one hand or the other as the drawing hand but not both together because that will arouse a conflict in our mind. This (first level) state of mind is here the referrer while the hands are the referent, but then we may observe, from a higher state of mind, both hands as drawing each other. Here this higher state of mind is the referrer level, while the separate observations of the single hands are in the referent level.

Figures \ref{fig:hands} \& \ref{fig:cupfaces} are two examples of a variety of drawings commonly known as `optical illusions', but this title ia misleading. Observing these drawings may be considered as performing experiments with the research question \emph{`What do we see?'}, and all these drawings are simply manifestations of how raw data -- a set of pixels in these cases -- may be interpreted in our mind in different ways depending on what we choose to focus upon.

The drawings deliver, therefore, a very important insight -- while we collect in our observations raw physical data, the same phenomenon may be interpreted in our mind in different modes. The distinction occurs in our mind.

Moreover, these examples explicitly demonstrate that we may observe ourselves making these observations and interpretations -- as in observing ourselves changing focus from one hand to the other and vice versa, or changing focus from cup to faces and vice versa. We then experience self-observation -- observation of an observation, with the lower level observation becoming an observable itself.

\vskip15pt

The graphical examples were brought to deliver the basic understandings regarding referencing observations without alluding initially to observations in physics, whose interpretations may be conrtroversial. We may now proceed formally as follows.

Let $\mathcal{X}$ be the space, realm, of all physical observables -- all the physical phenomena in the whole universe, from sub-atomic particles to clusters of galaxies, including all the processes and interactions therein, animate and inanimate. The result of an observation depends upon the point of view (\emph{e.g.}, a reference frame) from which we choose to perform the observation, and in many instances also upon the way we choose to interpret it. Let $a$ denote a typical point of view from which we observe such phenomena, and let $\mathcal{A}$ be the sum-total of all the possible points of view, $\mathcal{A} = \left\{a\right\}$. An observation is an act from some point of view $a \in \mathcal{A}$ to some physical phenomenon $x \in \mathcal{X}$ -- collecting some raw physical data, then referring to the image of the subject of observation in the observer's mind. This reference relation we denote in the following $a \searrow x$. A physics theory is therefore a collection of statements about the totality of our observations of the physical world, $\left\{a \searrow x\right\}$. We may use in the following the short-hand notation $\left\{\mathcal{A} \searrow \mathcal{X}\right\}$ for `all the observations $a \searrow x$ for all possible viewpoints $a\in\mathcal{A}$ and all possible physical phenomena $x\in\mathcal{X}$'.

With the observer being also part of the universe, the act of observation becomes an observable phenomenon. As demonstrated above with Figures \ref{fig:hands} \& \ref{fig:cupfaces}, observers (necessarily human) have the ability to observe their own process of observation. From any level of observation it is possible to observe only lower-order observations, not its own observations. Thus observers can develop another, \emph{higher} or \emph{superior}, point of view, $b$, from which they observe $a \searrow x$. So now we also have $b \searrow \left( a \searrow x \right)$, an observation of an observation, with $\mathcal{B} = \left\{ b \searrow \left( a \searrow x \right) | a \in \mathcal{A},  x \in \mathcal{X} \right\}$ the totality of \emph{observations of} $\mathcal{A}$\emph{-level observations}. Level $\mathcal{A}$ is referrer level for $\mathcal{X}$, but referent level for $\mathcal{B}$.

Hence, for Escher's ``Drawing hands'' (\Fref{fig:hands}), while $\mathcal{A}$-level viewing sees each of the hands \emph{either} as drawing \emph{or} as being drawn, $\mathcal{B}$-level viewing sees \emph{both} hands together, drawing \emph{and} being drawn. Let $a_1$ be \emph{viewing the right hand drawing the left hand} and $a_2$ be \emph{viewing the left hand drawing the right hand}. Then, with $x$ being the drawing itself, the $\mathcal{B}$-level viewing is $b \searrow \left\{ a_1 \searrow x , a_2 \searrow x \right\}$.

Similarly, for ``Rubin's vase'' (\Fref{fig:cupfaces}), with $a_1$ being \emph{focusing at the white area (cup)} and $a_2$ being \emph{focusing at the black area (faces)}, and $x$ being the drawing itself, $\mathcal{A}$-level viewings $\left\{ a_1 \searrow x , a_2 \searrow x \right\}$ are seeing \emph{either} the cup \emph{or} the faces, while $\mathcal{B}$-level viewing $b \searrow \left\{ a_1 \searrow x , a_2 \searrow x \right\}$ is seeing \emph{both} viewings (monitoring the possible switching of our focusing between the two).

Another example is the well-known story of the elephant in the village of blind people: The elephant itself is the subject $x$. It is the referent for the blind persons, each touching a different member of the elephant's body, being at different positions in level $\mathcal{A}$ which for them is the referrer level. But the wise man, who understands that they have touched different members of the same elephant, he is in level $\mathcal{B}$ which is the referrer level relative to the villagers.

\vskip30pt

\section{Referencing observations in physics}\label{sec: refobphys}

Classical physics is associated with the view that the human is only a witness in the evolution of the universe, and the observers are no more than passive bystanders. This is a very limited view, that ignores the fact the we -- humans, therefore the observers -- are integral part of the universe, together with our observations.

A characteristic of the classical level is that it can only accept that a physical system can be in only one, definite, state, defined by the detecting devices. Classical physics cannot accept co-existence of two or more different states simultaneously, let alone contradicting ones. Classical level observations are confined to observing, \emph{e.g.}, \emph{either} wave-like properties \emph{or} particle-like properties, according to the experimental setup and the properties of the detecting devices, but not both at the same time; or that an electron can pass \emph{either} through one slit \emph{or} through the other, but not both simultaneously.

In the graphical examples above, this corresponds to viewing each of the hands in \Fref{fig:hands} \emph{either} as drawing \emph{or} as being drawn, or to viewing \emph{either} the cup \emph{or} the faces in \Fref{fig:cupfaces}.

Classical physics corresponds therefore to the base level of observation -- direct observations of physical phenomena, $\left\{ \mathcal{A} \searrow \mathcal{X} \right\}$ in the notation above. Quantum mechanics, which suggests that in certain cases the same physical phenomenon may be observed differently in distinct modes, requires a higher-than-$\mathcal{A}$ level of observations.

Consider, for instance, the wave-particle duality. Let $x$ be certain EM radiation, with $a_1 \searrow x$ denoting observing its wave-like properties and $a_2 \searrow x$ denoting observing its particle-like properties. Classical, $\mathcal{A}$-level observations, are confined to observing \emph{either} the wave-like properties $a_1 \searrow x$ \emph{or} the particle-like properties $a_2 \searrow x$, because these are the detectable observables. But  quantum delayed-choice experiments indicate, quite convincingly, that EM radiation and photons are `wave-particle' -- an entity that combines both properties but is neither this nor that exclusively -- and it is only the nature of the detection device that exposes either the wave or particle aspect \cite{MaKofZeil16}. Therefore, as with the exemplary drawings above, quantum mechanics suggests a $\mathcal{B}$-level viewing $b \searrow \left\{ a_1 \searrow x , a_2 \searrow x \right\}$ needed to be able to appreciate that the EM radiation can be simultaneously in two distinct (classical) states. The understanding that we can arbitrarily choose between detecting the photon as a particle and detecting it as a wave is a $\mathcal{B}$-level appreciation.

In conclusion, a $\mathcal{B}$-level -- higher-than-$\mathcal{A}$ -- observation is needed to be able to appreciate that a physical entity can be simultaneously in more than one state (in the classical sense). In fact, the classically educated part of our mind, confined to $\mathcal{A}$-level observations only, cannot accept \emph{both} $a_1 \searrow x$ \emph{and} $a_2 \searrow x$ simultaneously. While $\mathcal{A}$-level observations ($\mathcal{A} \searrow x$) refer directly to some detectable physical phenomenon $x$, the subjects of $\mathcal{B}$-level observations are the $\mathcal{A} \searrow x$ observations themselves. Then the base level $\mathcal{A} \searrow \mathcal{X}$ observations become also observables, and higher $\mathcal{B}$-level observations $\mathcal{B} \searrow (\mathcal{A} \searrow  \mathcal{X})$ are necessarily involved.

Referencing observations appear in physics not only regarding quantum phenomena or experiments. The principle of relativity, which allows to relate measurements and observations from different points of view, therefore employing observations of observations, is necessarily at least  $\mathcal{B}$-level. Also, any complex or extended system can be viewed either as a whole, or just as collection of constituents. Then, as in the story of the elephant in the village of blind people, regarding it just as a collection of constituents is $\mathcal{A} \searrow \mathcal{X}$-level observation, corresponding to the blind villagers each touching a different member of the elephant's body, while appreciatining a whole-system view is already observing observations, therefore $\mathcal{B} \searrow (\mathcal{A} \searrow  \mathcal{X})$-level observations.

Yet another aspect of physics where referencing observations appear is the issue of time-asymmetry. Classical physics is time-symmetric, unable to account for the arrow of time, neither in mechanics nor in electrodynamics or cosmology. Time-asymmetry is introduced into classical physics by hand, as by assuming retardation or coarse-graining. Time-asymmetry observations are therefore at least $\mathcal{B}$-level.

\vskip30pt

\section{Hierarchical referencing and physics incompleteness}\label{sec: hieref}

$\mathcal{A}$-level observations, of directly physically detectable phenomena, gave rise to the classical theories of physics, mainly Newtonian mechanics and classical electrodynamics, with the principles of energy and momenta conservation. But an observation is also an observable phenomenon, and empirical evidence, unexplainable by classical physics -- such as the wave-particle duality, particles' two-slit interfrence, time-asymmetry, \emph{etc.} -- necessarily introduced questions regarding $\mathcal{A}$-level observations. These are therefore $\mathcal{B}$-level observations -- observations of $\mathcal{A}$-level observations, of the kind denoted earlier $\mathcal{B} \searrow (\mathcal{A} \searrow \mathcal{X})$.

There may be more than two reference levels of observation: From any level of observation it is possible to observe only lower-order observations, not its own observations. Hence, from $\mathcal{B}$ it is possible to observe $\mathcal{A} \searrow \mathcal{X}$, \emph{i.e.}, $\mathcal{B} \searrow \left(\mathcal{A} \searrow \mathcal{X} \right)$ is possible, but it cannot be observed from $\mathcal{B}$. Still, the observer may develop a higher-order level of view $\mathcal{C}$ from which it is possible to observe $\mathcal{B} \searrow \left(\mathcal{A} \searrow \mathcal{X} \right)$; namely, $\mathcal{C} \searrow \left(\mathcal{B} \searrow \left(\mathcal{A} \searrow \mathcal{X} \right)\right)$ is possible, \emph{etc.}.

Indeed, recent delayed-choice experiments reviewed in \cite{MaKofZeil16} indicate that even the distinction between `entanglement' and `separability' depends on the detecting device and is not an essential characteristic of a physical system. This distinction is an appreciation of $\mathcal{B}$-level observations, therefore has to be a $\mathcal{C}$-level observation.

While these $\mathcal{B}$-level observations are recognized empirically, and certainly appear in discussions and dialogues among physicists, they are not accounted for, so far, theoretically. Physics theories don't know yet how to include observers and observations and how to refer to participating observations $ \mathcal{A} \searrow \mathcal{X}$ as part of the subject matter\footnote{A two-stage quantum measurement was proposed by Zwick \cite{Zwick78}. He ignores, though, the r\^{o}le of the observer in the process.}. The higher B-level observations $\left\{\mathcal{B} \searrow \left(\mathcal{A} \searrow \mathcal{X}\right)\right\}$ call for insights and understandings above and beyond the lower $\mathcal{A}$-level observations $\left\{ \mathcal{A} \searrow \mathcal{X} \right\}$, and these have to be part of a new, futuristic theory.

Similarly, new empirical evidence regarding $\mathcal{B}$-level observations, such as the non-essentiality of the distinction between `entanglement' and `separability', call for new insights and understandings above and beyond the $\mathcal{B}$-level understandings, which are then contained in $\mathcal{C}$-level understandings.

Recalling that observation is a unidirectional act, from observer to observed, so that from an observation level it is possible to observe lower-order observation levels, we distinguish between the referrer (observer) and referent (observed) levels, putting the former at superiority relative to the latter. Hence, in observing the universe we construct hierarchies of observation levels (points or levels of view):
\begin{itemize}
  \item {1$^{\rm st}$ order observation: Direct observations of physical phenomena, $\left\{\mathcal{A} \searrow \mathcal{X}\right\}$.}
  \item {2$^{\rm nd}$ order observation: Observing 1$^{\rm st}$ order observations, $\left\{\mathcal{B} \searrow \left(\mathcal{A} \searrow \mathcal{X}\right)\right\}$.}
  \item {3$^{\rm rd}$ order observation: Observing 2$^{\rm nd}$ order observations, $\left\{\mathcal{C} \searrow \left(\mathcal{B} \searrow \left(\mathcal{A} \searrow \mathcal{X}\right)\right)\right\}$.}
  \item {\emph{etc.}}
\end{itemize}

In simile, such a hierarchy is found in the structure of the universe, from micro to macro, as in
$$\rm{particles} \rightarrow \rm{atoms} \rightarrow \rm{molecules} \rightarrow \rm{crystals} \rightarrow \rm{planets} \rightarrow \rm{galaxies} \rightarrow \ldots$$
or
$$\rm{atoms} \rightarrow \rm{molecules} \rightarrow \rm{cells} \rightarrow \rm{organizms} \rightarrow \ldots$$
Viewing each level is focusing our mind on that level.

When a new phenomenon is observed that the current theory cannot explain, we seek for new first principles that cover these new observations. Therefore, in principle at least, this process of new higher level observations referring to lower level observations can go on and on (or up and up), without limit. Together they create a hierarchy of observations, which may go on endlessly.

Each level in the hierarchy of observation levels necessarily contains insights and understandings regarding the lower levels. With each level adding some insights and understandings that are not contained in the levels below, the unlimited hierarchy of observation levels cannot be summed up with a finite number of insights and understandings -- thus inevitably implying incompleteness.

Indeed, in G\"{o}del's proof, for any finite set of first principles and inference rules there exist some statements which are undecidable due to paradoxical self-negation. These statements may become provable (or refutable) by adding new axioms, or first principles, but then new undecidable statements ensue\cite{Raatikainen}. A finite set of axioms will never be enough to make the theory complete -- a never-ending series, or hierarchy, of new axioms is required, which is essentially the incompleteness. By such a hierarchy the paradoxical self-negation is circumvented, but the incompleteness remains unavoidable.

\vskip15pt

Self-reference appears when we are not aware of the distinction between levels of observation. In a way, it is like looking at a photograph of some panoramic landscape. By taking the photo we actually squeeze a 3D source into a 2D result. The photograph doesn't distinguish between different depths -- in the photo they are all the same level. It is only that we, with our experience and knowledge of the laws of perspective, distinguish between close and distant when we look at the photograph. This distinction occurs wholly in our minds -- there we re-discover the depths again.

It is the same with any image that we form in our mind regarding the universe -- we can see it squeezed, no hierarchical levels, or, alternatively, we can see and identify the hierarchies and the levels of which they are constructed.

Therefore, when the distinction between levels is ignored, we get a picture in which everything -- the contents of all the levels -- is squeezed into just one common level. This naive situation is illustrated in \Fref{fig:supersystem}. Then we encounter self-referencing, and logical paradoxes may occur. In other words, self-reference is an indication for squeezing separate levels into one. Paradoxes are indications of misconceptions, and self-reference paradoxes appear when levels that should be separated are instead squeezed (in our mind) into just one common level.

\vskip30pt

\section{Final notes}\label{sec: ff}

The results of G\"{o}del's theorem are used here as a lead to consider, in a sense, the future of physics theories. A universal point of view, beyond current physics, is proposed :

We (humans) are part of the universe. Physics theories are formed of the interpretatios that we make of what we observe in the universe.

By our daily decisions, in the way we act and observe our actions, we take part in determining how the universe evolves. In particlar, by choosing the mode of observation in certain experiments we determine the nature of the outcome.

An observation is therefore also an observable phenomenon, we are both observers and observed, and we and our observations should also be considered part of the subject matter of physics theories.

A universal physics theory will ever be incomplete. First, for the very simple reason that no one can assure us that new facts, that require drastic changes in how we view the physical world, will never be discovered sometime in the future. New discoveries imply new first principles in the foundation of the theory, new insights that are not derivable from old ones. The scientific research will then produce, in the way it used to be and what seems to be a never-ending process, more and more insights, understandings and knowledge, within larger and larger theories.

Second, because our observations introduce referencing into physics. The foregoing discussion indicates that since we may observe our own observations, and from any level of observation it is possible to observe only lower-order observations, then the various observations must form some hierarchy. We interpret the world according to the results of our observations. Levels of observations imply levels of interpretations. Each level implies a more profound insight, associated with an essentially new discovery. Therefore, (at least in principle) an endless hierarchy of levels of interpretation.

We thus conclude with another quotation of Hawking:
\begin{quote}
\textit{Some people will be very disappointed if there is not an ultimate theory that can be formulated as a finite number of principles. I used to belong to that camp, but I have changed my mind. I'm now glad that our search for understanding will never come to an end, and that we will always have the challenge of new discovery. Without it, we would stagnate. G\"{o}del’s theorem ensured there would always be a job .. for physicists.} \cite{Hawking02}
\end{quote}

The next fundamental change in physics theories will come -- and here the place for my speculations --  when we know how to include, within the framework of theory, the observer as an integral participant in physics events. J.A. Wheeler suggested that our investigations and inquiries determine, at least in a way, how the universe develops. Religeous traditions, like the Jewish, maintain that the human is an active participant in the evolution of the universe. Could this be an indication for a new generation of physics theories?

\vskip30pt

\noindent{\textbf{Acknowledgements.} I wish to thank Noson Yanofsky for fruitful communications.}

\vskip30pt

\rule{10cm}{1pt}

%References:

\end{document}